\documentclass{eas}
\usepackage{graphicx}
\begin{document}

\title{Theory and Models of the Disc-Halo Connection}
\author{D. Breitschwerdt}\address{Zentrum f\"ur Astronomie und Astrophysik, Technische Universit\"at Berlin, Hardenbergstr.~36, D-10623 Berlin, Germany}
\author{M.A. de Avillez}\address{Department of Mathematics, University of \'Evora, R. Rom\~ao Ramalho 59, 7000 \'Evora, Portugal}
\author{V. Baumgartner}\address{Institut f\"ur Astronomie, University of Vienna, T\"urkenschanzstr.~17, A-1180 Vienna, Austria}
\author{V.A. Dogiel}\address{P.N. Lebedev Institute, Leninskii
pr, 53, 119991 Moscow, Russia}
\begin{abstract}
We review the evolution of the interstellar medium in disc galaxies, and show, both analytically and by numerical 3D hydrodynamic simulations, that the disc-halo connection is an essential ingredient in understanding the evolution of star forming galaxies. Depending on the star formation rate of the underlying gaseous disc, a galactic fountain is established. If the star formation rate is sufficiently high and/or cosmic rays are well coupled to the thermal plasma, a galactic wind will be formed and lead to a secular mass loss of the galaxy. Such a wind leaves a unique imprint on the soft X-ray spectra in edge-on galaxies, with delayed recombination being one of its distinctive features. We argue that synthetic spectra, obtained from self-consistent dynamical and thermal modelling of a galactic outflow, should be treated on an equal footing as observed spectra. We show that it is thus possible to successfully fit the spectrum of the starburst galaxy NGC$\,$3079.
\end{abstract}
\maketitle
\section{Introduction}
\label{intro}
%
Historically, Pikel'ner (1953) inferred the existence of a ``rarefied gas'' distributed in a ``spherical subhalo'' around the Galaxy, by arguing that the confinement of cosmic rays (CRs) by a magnetic field required a large velocity spread in the dilute medium between clouds, so that its kinetic energy was of the same order as the magnetic energy, leading to the extension of the gas perpendicular to the disc. Somewhat later, Spitzer (1956) has postulated the existence of a hot galactic corona in order to confine extraplanar high velocity clouds by external pressure. ROSAT PSPC observations have confirmed the existence of X-ray emitting halos for a number of late type edge-on galaxies like e.g.~NGC$\,$891 (e.g.~Bregman and Houck 1997), NGC$\,$4631 (e.g.~Wang et al.\ 2001). Recently, among others, T\"ullmann et al.\ (2006) have demonstrated for a sample of FIR selected galaxies that there exists a link between star formation rate (SFR) and extended X-ray halos. Physically, for typical numbers of the halo gas density, $n_e = 10^{-2} \, {\rm cm}^{-3}$, and temperature, $T=10^6 \, {\rm K}$, of the Milky Way, the sound crossing time scale over a vertical extension of 10 kpc is larger than the cooling time, i.e., $\tau_{\rm dyn} \gg \tau_{\rm cool}$. Since $\tau_{\rm dyn} \sim c_s/g_z$, where $c_s$ and $g_z \approx 10^{-9} \, {\rm cm}/{\rm s}^2$ are the sound speed and the vertical component of the gravitational acceleration, respectively, and $\tau_{\rm cool} \sim P^{3/2}/(q \rho^{5/2})$, with $P, \rho$ being the gas pressure and density, respectively, and $q= 4 \times 10^{32} \, {\rm cm}^6 \, {\rm g}^{-1} \, {\rm s}^{-4}$, the condition for a hydrostatic corona translates according to Kahn (1981) into $P/\rho^2 \gg q/g_z$, which is not met by the numbers given above. Therefore, the hot supernova (SN) shock heated gas streaming from the disc into the halo will in general not attain hydrostatic equilibrium, unless its metallicity is very low (thereby increasing its cooling time). This has been realized early on and has led to the concept of the galactic fountain (Bregman 1980, Kahn 1981) and wind (Breitschwerdt et al.\ 1991) models. The fountain model has been originally suggested by Shapiro and Field (1976), who tried to explain both the observed soft X-ray background (Williamson et al.\ \cite{wil74}) and the recently discovered interstellar O{\sc vi} absorption line with Copernicus (Jenkins \& Meloy \cite{jm74}) by a single hydrostatic model, and found that it was virtually impossible to have a hot component in pressure equilibrium with the observed cooler gas. In addition, the measured O{\sc vi} line width was in some cases much narrower than would have been allowed by a gas with $T \ge 10^6 \,$K. The only reasonable conclusion was therefore to cool the gas by adiabatic expansion (as well as by line radiation) when it rises into the halo. The most efficient way to transport hot gas into the halo is by forming superbubbles (SBs) powered by supernova explosions correlated in time and space. In Section~2 we describe by  an analytical model how a superbubble expands in a density stratified medium and derive the conditions for blow-out. Part of the outflow will leave the galaxy as a wind, in particular if CRs assist in driving the flow (Breitschwerdt et al.\ 1991, Everett et al.\ 2008), which will be discussed in Section~3. It will also be shown that X-ray halos may bear the signature of such winds. Section~4 will close with a summary and conclusions.

\section{Superbubble Dynamics}
\subsection{Analytical Approach}
The majority of massive stars is found in associations rather than randomly distributed.
Winds and subsequent SNe create large holes in the HI layer of a galactic disc, as observed by Heiles (\cite{h84}). They are surrounded by the swept-up ISM, which collapses into thin, cool and dense shells soon after the formation of these SBs.
If the disc is stratified and with sufficiently large energy input, such a bubble will accelerate along the density gradient of the ISM, channeling the gas from the disc into the halo of a galaxy, which is often called a chimney. Once acceleration has started, the shell is susceptible to Rayleigh-Taylor instabilities and will break up in its further evolution. The hot, metal enriched gas is expelled into the halo or into the local intergalactic medium.
This blow-out phenomenon was studied in many numerical approaches (e.g.~Tomisaka \& Ikeuchi \cite{ti86}) or by a combination of both numerical and analytical models (e.g.~Mac Low \& McCray \cite{mm88}).
In its initial stages, the evolution of a SB can be described as if expanding spherically into a uniform medium (Weaver et al.\ \cite{w77}). At later times, when the size exceeds a density scale height, this assumption is not valid anymore. Analytically, the approximation by Kompaneets (Kompaneets \cite{k60}) can be applied to the evolution of SBs in an exponentially stratified disc.
It is assumed that a point-like explosion takes place in a purely exponential atmosphere, but can be modified to include a continuous energy input (e.g.~Basu et al.\ \cite{bjm99}) or -- as it is done in our analytical approach --
a time-dependent energy input rate given by an Initial Mass Function (IMF) for cluster stars (Baumgartner \cite{ba07}).
With this model, we find the analytical expressions for velocity and acceleration of the outer shock, as the bubble expands in a density stratified medium given by $\rho(z)= \rho_0 \cdot \rm{exp}(-z/H)$ with $\rho_0$ being the midplane number density.
Moreover, we are able to give the number of stars that is needed for blow-out as a function of scale height and density of the ambient medium.

Several assumptions are made in Kompaneets' model, which are reviewed here briefly:
(i) A strong shock is propagating into the undisturbed ISM, (ii) the mass is concentrated in a thin shell,
(iii) there is uniform pressure inside the bubble. The explosion energy, thermalized at the inner shock, is
$
E_{\rm{th}} = \frac{1}{\gamma - 1} \cdot P(t) \cdot V(t)
$
in a bubble with volume $V(t)$ and adiabatic interior with pressure $P(t)$. We use a ratio of specific heats of $\gamma =5/3$. By using a transformed time variable (e.g.~Bisnovatyi-Kogan \& Silich \cite{bks95})
$
y=\int_{0}^{t} \sqrt{\frac{\gamma^2 - 1}{2} \frac{ E_{\rm{th}} (t)}{\rho_0 \cdot V(t)}} \, dt
$
the half-width extension of the bubble parallel to the galactic plane is obtained
\begin{equation}
r(y, z)=2H\arccos \left[ \frac{1}{2} e^{z/2H} \left(1 - \frac{y^2}{4H^2} + e^{-z/H} \right) \right].
\label{V1}
\end{equation}
%
From the solution we see that top and bottom of the bubble are given by $z_u(\tilde{y}) = -2H \ln\ (1-\tilde{y}/2)$ and $z_d(\tilde{y}) = -2H \ln\ (1+\tilde{y}/2)$, respectively, with $\tilde{y} = y/H$.
%
%
%
%
%
%
To proceed with our analytical work we simply describe the volume of the bubble
by an ellipsoid
$
V (\tilde{y}) = \frac{4\pi}{3}a(\tilde{y})r_{\rm{max}}^2(\tilde{y}) \,,
$
%
with a semimajor axis $a(\tilde{y}) = (z_u(\tilde{y}) - z_d(\tilde{y}))/2$ and the semiminor axis corresponding to the maximum half-width extension of the bubble in $r$-direction $r_{\rm{max}}=2H \arcsin\ (\tilde{y}/2)$.
This approximation is very good even at late evolutionary stages of a bubble (with errors of less than
1 $\%$ at $\tilde{y} = 1.7$).
%
%
The sequence of exploding stars is modelled using an IMF and the main sequence life time of massive stars.
The IMF gives the probability to form a number of stars $dN/dm = N_0 \cdot m^{\bar{\gamma}}$ in a mass interval $(m_l, m_u)$.
$N_0$ is a normalization factor, which can be obtained by fixing the number of stars in the last mass bin $N(m_u-1, m_u)$ to be equal to one (Bergh\"ofer \& Breitschwerdt \cite{bb2002}).
We choose an IMF with slope $\Gamma = \bar{\gamma} + 1 = -1.35$ (Salpeter \cite{sp55}).
In order to relate the mass of a star to a time scale, we use the main sequence life time derived by Fuchs et al.\ (\cite{f2006})
$\tau_{\rm ms} = \tau_0 \, M^{-\eta}$, with $\tau_0=1.6 \times 10^8$ yr and $\eta =0.932$, for $2 < M < 67$ (with $M$ in units of M$_{\odot}$), such that the result is a time-dependent energy input rate (e.g.~Bergh\"ofer \& Breitschwerdt \cite{bb2002}; Fuchs et al.\ \cite{f2006}). It is given by $L_{\rm{SB}} = L_{\rm{IMF}} \cdot t^{\delta}$
with $L_{\rm{IMF}} = \left(E_{\rm{SN}} \cdot N_0 \cdot (\tau_0)^{\Gamma/\eta}\right)/\eta$
and $\delta = - (\Gamma/\eta + 1)$. We obtain a characteristic time scale for our model
$t_{\rm{IMF}}= (\rho_0 \cdot H^5/L_{\rm{IMF}})^{1/(\delta +3)}$.
Similar to the Weaver et al.\ (1977) solution for a bubble blown by continuous energy input in a homogeneous medium, we derive, by implementing our new time-dependent energy input rate, the thermal energy inside the hot plasma filling the bubble, $E_{\rm{th}} = \frac{5}{7 \delta +11} \cdot L_{\rm{IMF}} \cdot t^{{\delta}+1}$.
We are thus able to calculate velocity and acceleration of the bubble for all points of the surface, where we get an explicit
expression for the coordinate from the solution (see Eq.~\ref{V1}).
By searching for the minimum of the velocity, we find that blow-out (the change from deceleration to acceleration of the shock front at $z_u$) occurs at $\tilde{y}_{\rm{acc}}\cong0.68$, or, given in units of the characteristic time scale, at $t_{\rm{blow}}=t(\tilde{y}_{\rm{acc}})/t_{\rm{IMF}} \cong 0.99$.
\begin{figure*}
   \centering
\hspace{-0.15\textwidth}
  \begin{minipage}{0.4\textwidth}
	\includegraphics[width=1\textwidth, angle=270]{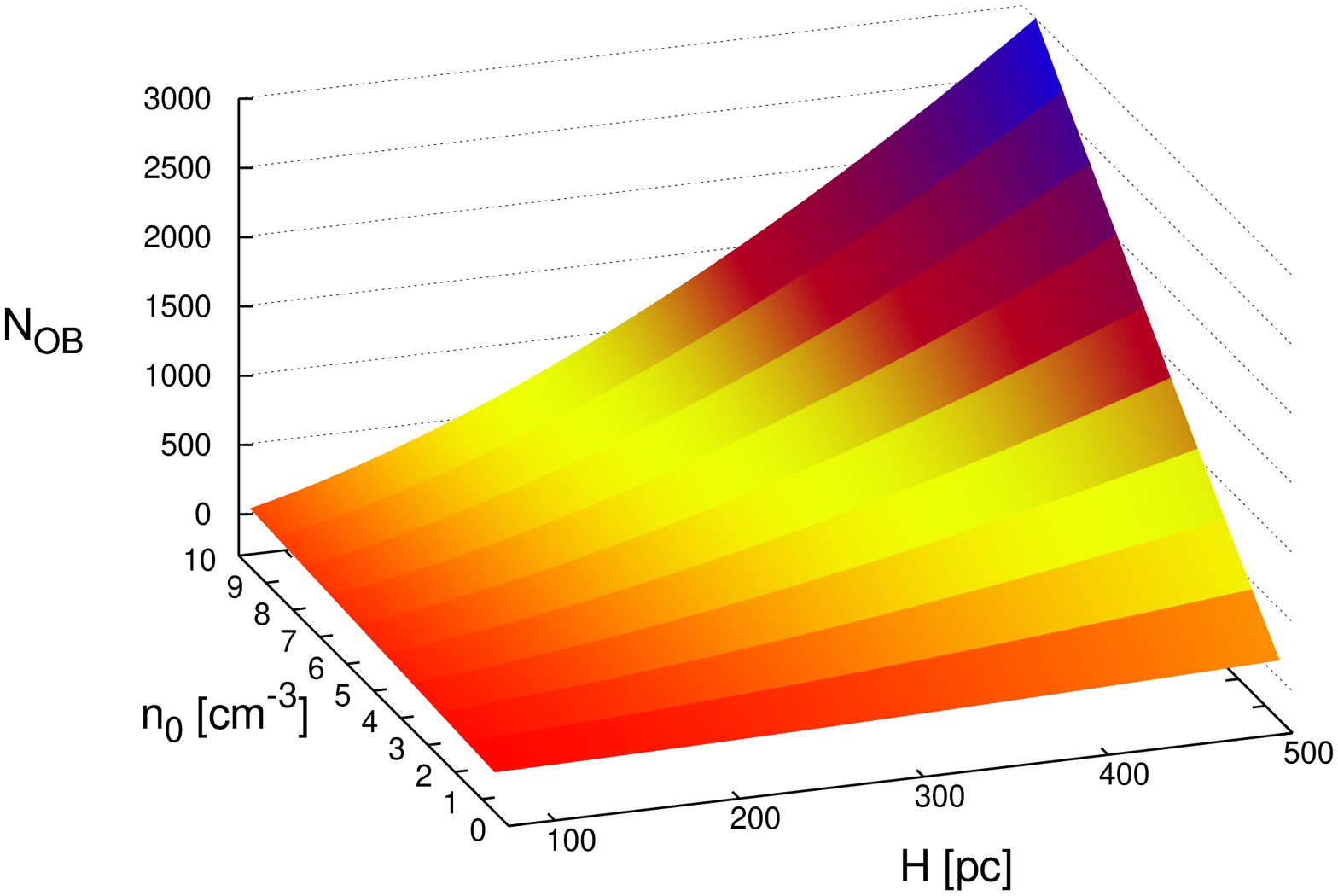}
  \end{minipage}
  \hspace{0.1\textwidth}
  \begin{minipage}{0.325\textwidth}
	\includegraphics[width=1\textwidth, angle=270]{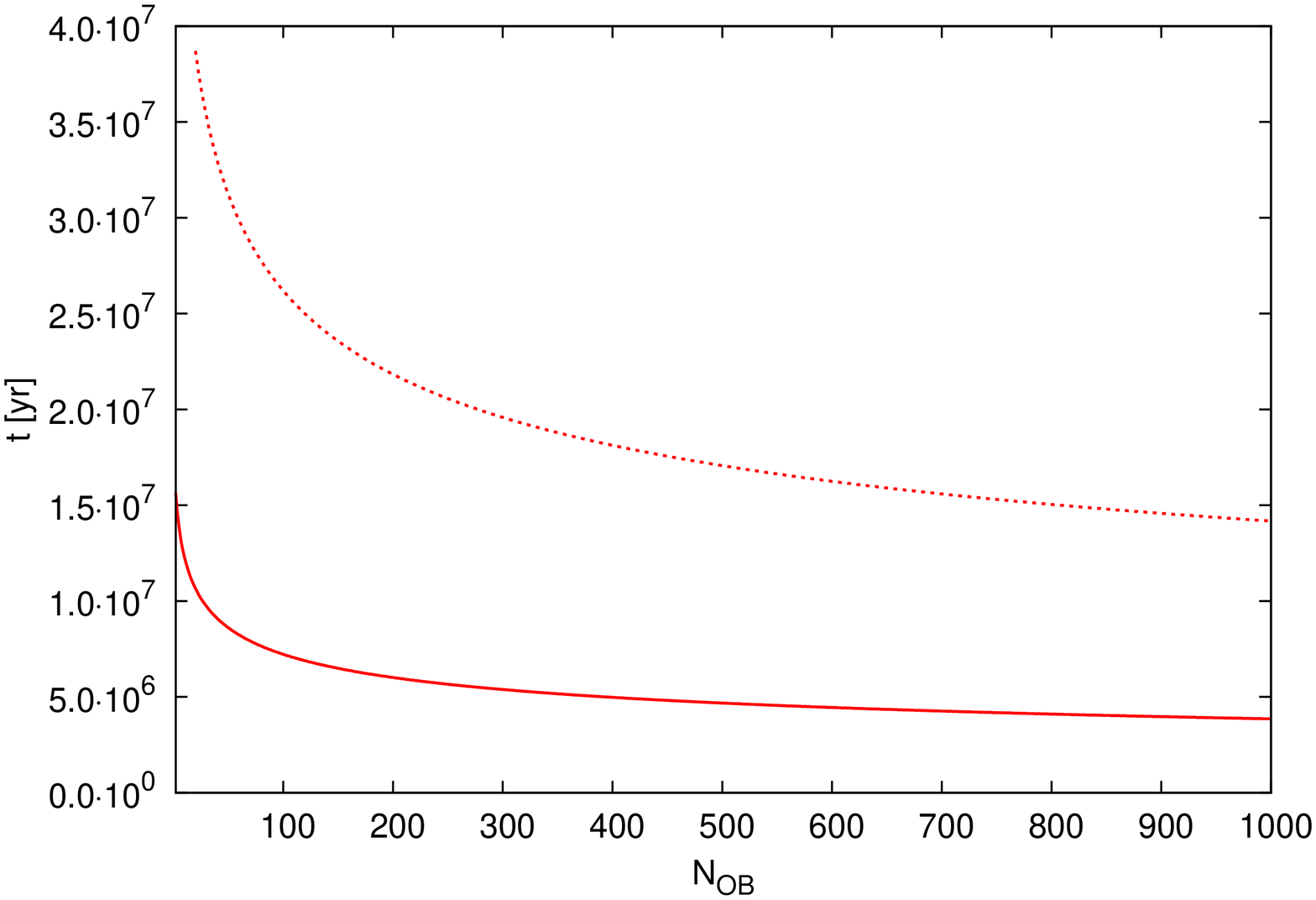}
  \end{minipage}
\caption{\textit{Left:} The number of SNe, ${\rm N}_{\rm OB}$, needed for blow-out for different combinations of scale height, ${\rm H}$, and number density, $n_0$, of the surrounding ISM.
\textit{Right:} Blow-out time scale (solid line) and fragmentation time scale (dotted line) at the top of the shell as a function of OB-association member stars.}
\hspace{0.0\textwidth}
\label{blowout}
\end{figure*}
By fixing the value of $\tilde{y}$, where the acceleration starts, and assuming that a
Mach number of $M \ge 5$ ($\sim30\,$km/s) justifies the assumption of a strong shock, we derive the number of stars required for blow-out as a function of ISM parameters (see Fig.~\ref{blowout}, left). For the Lockman layer of warm neutral gas (Lockman \cite{lock84}) with a scale height of $H \sim 500$\,pc, a number density of $0.1\,\rm{cm}^{-3}$ and a temperature of $\sim6000$\,K, we get a minimum number of roughly 20 stars.
Fig.~\ref{blowout} (right, solid line) shows the blow-out time scale into the Lockman layer for clusters with different richness. A SB produced by a cluster of 20 stars accelerates about 10~Myr after the first SN exploded.
Shortly after the acceleration of the outer shock sets in, instabilities start to appear at the top of the bubble shell on a time scale
\begin{equation}
t_{\rm{RTI},z_u}= \sqrt{ \frac{\Delta d(\tilde{y})}{2 \pi \ddot{z}_u(\tilde{y})} \cdot \frac{4 \rho_0 e^{-z_u(\tilde{y})/H} + \rho_{\rm{in}}(\tilde{y})} {4 \rho_0 e^{-z_u(\tilde{y})/H} - \rho_{\rm{in}} (\tilde{y})} }
\label{V5}
\end{equation}
The perturbation wavelength is of the size of the shell thickness $\Delta d(\tilde{y})$, the density in the bubble interior is given by $\rho_{\rm{in}} (\tilde{y})$ and $\ddot{z}_u(\tilde{y})$ denotes the acceleration at this coordinate of the bubble.
The instabilities start to dominate, when this time scale becomes smaller than the dynamical time scale of the SB, i.e., $t_{\rm{dyn}} = \frac{R}{\dot{R}} = \frac{z_u}{\dot{z}_u}$, which happens at $\tilde{y}_{\rm{RTI}} \cong 0.69$.
We assume that the instability is fully developed after about three times $t_{\rm{RTI},z_u}$.
Fig.~\ref{blowout} (right, dotted line) shows how this fragmentation time scale depends on the number of OB-stars.
It has to occur in a reasonable time scale of about 40~Myr, not long after the least massive star exploded and before the bubble may be distorted by shear motions or turbulence in the ISM, and thus we search again for the minimum number of cluster members to fulfill this criterion. The result corresponds very well to the number found for blow-out of 20 stars. In this case, the top of the bubble is broken up after $\sim\,$50~Myr and hot, metal enriched material is expelled to the high-$z$ regions of the halo.
\subsection{Numerical Simulations}
The ISM in star forming galaxies is a highly compressible turbulent medium. The Reynolds number is much higher than achievable in any terrestrial laboratory. This is mainly due to high velocities and large scales involved, which are particularly perceivable by the effects of supernova activity. Hence a model with a minimum degree of realism should be a numerical one, since nonlinear waves and mode coupling is difficult to grasp analytically. During the last decade, progress in 3D parallel computing and cheap hardware made it possible to tackle the problem. However, many assumptions, which had to be made (e.g.~small computational boxes, short evolution time scales, treating only small patches of ISM, applying artificial boundary conditions, lack of suitable closure models etc.), and often neglecting essential physics, show that we still have a long way to go. Nevertheless, the results, emphasizing different aspects of the ISM evolution, are intriguing (cf.\ Korpi et al.\ 1999, Wada and Norman 1999, Joung and MacLow 2006, Gressel et al.\ 2008), and give insight into the physical processes, which are at work. Our numerical 3D high resolution simulations (Avillez and Breitschwerdt 2004, 2005, henceforth AB04, AB05, respectively) focus on long evolution times and sufficient spatial resolution to track cooling and small scale instabilities.

The 3D \'Evora-Vienna Astrophysics Fluid-Parallel AMR (EVAF-PAMR) Code, which is based on the code by Avillez (2000), follows the detailed evolution of a Galactic patch up to 400 Myr in time, starting from a gas distribution as observed at present. The gaseous disc is immersed in a stellar gravitational field as described by Kuijken \&
Gilmore (1989). Other physical processes implemented in the code are (i) local self-gravity (excluding the contribution
from the newly formed stars), (ii) radiative cooling, assuming collisional ionization
equilibrium (following Dalgarno \& McCray (1972) and Sutherland \& Dopita (1993)
cooling functions for gas with $10 \leq T < 10^{4}$ and $10^{4} \leq T
\leq 10^{8.5}$ K, respectively) using solar abundances (Anders \& Grevesse 1989),
(iii) uniform heating due to a UV radiation field normalized to the Galactic value and
varying with $z$ (cf. discussion in Wolfire et al.\ 1995). The disc gas is energized by
SNe types Ia and II (including types Ib and Ic) with the observed
Galactic SN rates (Cappellaro et al.\ 1999), using a finest resolution of 0.625 pc on a global grid extending 1 kpc in the disc in $x$- and $y$-directions and $\pm 10$ kpc in $z$-direction. The boundary conditions are chosen to be periodic along the vertical and outflow on the top and bottom surfaces of the grid.

\begin{figure*}
  \centering
  \includegraphics[width=0.6\hsize,angle=-90]{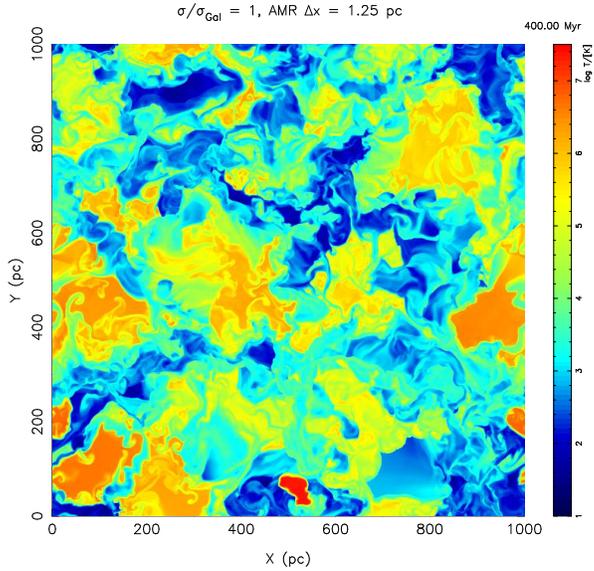}

\caption{Temperature distribution in a 2D cut through the 3D data cube in the Galactic midplane for a Galactic supernova rate. }
\label{disktemp1}       
\end{figure*}

Fig.~\ref{disktemp1} shows the colour coded temperature distribution in a cut through the Galactic midplane centred on the solar circle after an evolution time of 400 Myr. SNe occur both in single and in spatially and temporally correlated explosions, stirring up the ISM and heating it up to temperatures of $10^7 \,$K or more. The most striking feature of the interstellar plasma is the existence of structures at all resolvable scales, which is a result of strong turbulence. This can be understood in terms of generating shear and vorticity with subsequent compression and stretching of vortex tubes. A striking consequence of this behaviour is the existence of gas in classical thermally unstable phases (in contrast to the 3-phase model by McKee and Ostriker 1977, henceforth MO77), as it has been observed in warm neutral H{\sc i} gas (Heiles and Troland 2003). It turns out that turbulent diffusion has a stabilizing effect on gas in temperature regimes subjected to thermal instabilities. Moreover, the volume filling factor of the hot ($\geq 10^6 \,$K) gas is below 20\% and not $\geq 50$\% as predicted by MO77. This is due to the establishment of the galactic fountain cycle within a time scale of $\sim 200 \,$Myr. There have been arguments in the past (e.g.~Tomisaka 1990) that a disc parallel magnetic field would suppress blow-out, and hence the fountain. However, we have shown (AB05) in 3D high resolution MHD simulations, that this is not the case. One reason for this result  may be the fact, that in a SN driven ISM the disc is not homogeneous, and it is not permeated by a $z$-independent B-field, but the field is coupled to the density of the cold gas. The fountain in a spiral galaxy is far from homogeneous, and its energy source may vary in time with star formation rate. Therefore part of the outflow may have sufficient energy to leave the gravitational potential as a wind, in particular, if cosmic rays assist in driving the flow.
\section{Galactic Winds}
%
\begin{figure*}
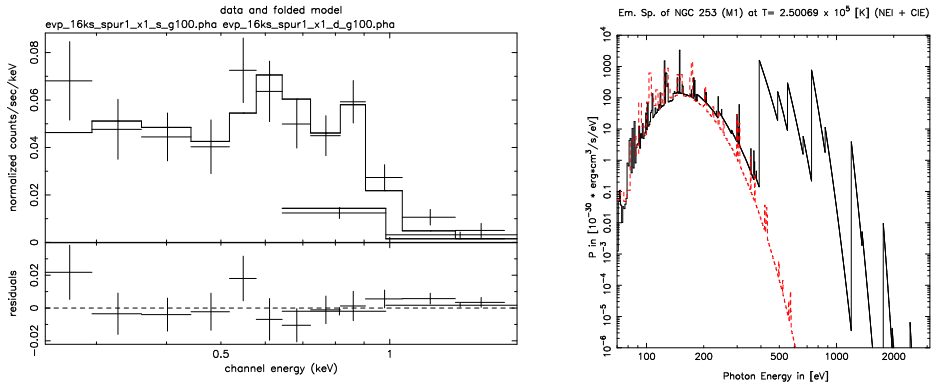

\centering
\includegraphics[width=0.4\hsize,bb=80 45 562 775,angle=-90,clip=]{n3079_sp_fit.ps}
\includegraphics[width=0.4\hsize,angle=-90,clip=]{n253_m1_comp_NEI-CIE_1.ps}
\caption{\emph{Left:} Thermally and dynamically self-consistent fit model (solid line) of NGC$\,$3079 (cf. Fig.~\ref{outflow}) of XMM-Newton observations (crosses). The bottom panel shows the residuals of the fit.  \emph{Right:} Simulated \emph{intrinsic} spectrum of the halo at a temperature of $2.5 \times 10^5 \,$K (dashed line: CIE model, solid line: NEI) for the dynamically and thermally self-consistent galactic wind solution (see Fig.~\ref{outflow}). Note that the recombination contributes up to energies of 2 keV(!) in the NEI case.
\label{spectra}
}
\end{figure*}
Cosmic rays (CRs) are the high energy component of the ISM with an energy density comparable to the thermal gas. If the coupling of CRs to the gas, via MHD waves, is strong, they will transfer momentum to the gas and thereby exert a pressure (for details see the paper by Dogiel and Breitschwerdt, this volume). Thus the gas will be dragged along as the CRs escape from the galaxy. Coupling will be facilitated by the streaming instability, which arises from a small scale anisotropy in the CR distribution function due to their net motion away from the galaxy, resulting in resonant generation of waves, which in turn scatter the particles strongly and thus ensure strong coupling. The theory of CR driven winds has been worked out in detail by Breitschwerdt et al.\ (1991, 1993) and Zirakashvili et al.\ (1996) for a steady-state outflow. In order to keep things simple, we have chosen a flux tube geometry, in which the area cross section makes a smooth transition from plane parallel to spherical geometry at the distance of a galactic radius, $R_g$, by $A(z) = A_0 (1+ z^2/R_g^2)$, with $z$ being the coordinate perpendicular to the disc and $A_0$ a surface element in the midplane. For a given pressure distribution $P_{g,0}$ (gas), $P_{c,0}$ (CRs), and $P_{w,0}$ (waves), and a specified gas density, $\rho_0$, the boundary condition of zero intergalactic pressure at infinity ensures the existence of a Parker type solution, for which the initial velocity, $u_0$, and hence the mass loss rate per unit area, $\dot m=\rho_0 \, u_0$, will be determined. It turns out that for large distances the velocity tends to an asymptotic value, $u_\infty$. Therefore due to mass conservation the density drops off with distance like $\rho(z) \propto 1/z^2$ rather than exponentially, as in a halo in hydrostatic equilibrium. The density signature can be tested by deep halo observations of edge-on galaxies in soft X-rays.

\subsection{Modelling Galactic Halos}
\begin{figure*}
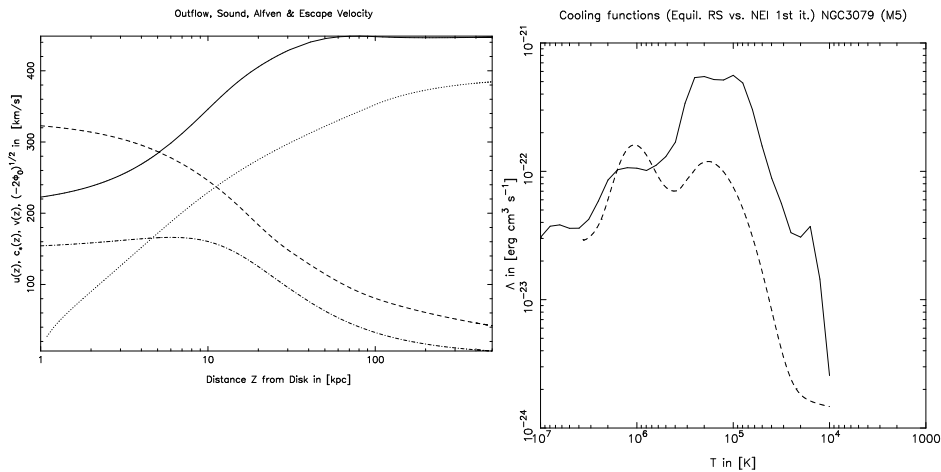

\centering
\includegraphics[width=0.4\hsize,angle=-90,clip=]{n3079_m5_vel.ps}
\includegraphics[width=0.49\hsize,angle=-90,clip=]{coolc_n3079_m5-3it.ps}
\caption{\emph{Left:} Galactic wind outflow solution (solid line: outflow velocity, $u$, long dashed line: compound sound speed, $c_*$, dashed-dotted line: Alfv\'en velocity, $v_{\rm A}$, dotted line: escape velocity as a function of distance) for a model reproducing the XMM-Newton spectrum of the north eastern halo of NGC$\,$3079. \emph{Right:} Cooling function (solid line: collisional ionization equilibrium (CIE), dashed line: non-equilibrium ionization (NEI)) for the dynamically and thermally self-consistent galactic wind solution (see left figure).
\label{outflow}
}
\end{figure*}
We have seen that outflows modify the halo structure (e.g.~density and temperature) considerably, which leaves a signature in associated soft X-ray spectra. Fig.~\ref{spectra} (left) shows as an example the spectrum of the north eastern part (``spur 1'') of the starburst galaxy NGC$\,$3079, observed with XMM-Newton. The fit (solid line) is the result of a dynamically and thermally self-consistent model as discussed by Breitschwerdt and Schmutzler (1999). In a nutshell, the modified density and temperature structure changes the amount of radiative cooling, since the cooling function is $\Lambda =\Lambda(\rho,T)$,  which in turn modifies the outflow and therefore the velocity and density structure, thus closing a nonlinear feedback loop (s.~Fig.~\ref{outflow}). For high star formation rates, as it is the case for starburst galaxies, the outflow is so strong that adiabatic dominates radiative cooling, and, in particular, the dynamical is much shorter than the recombination time scale. Consequently, the signature of such an overionized plasma is soft X-ray emission at temperatures, which are much too (s.~Fig.~\ref{spectra}, right) low in case of a gas in collisional ionization equilibrium (CIE). Whereas CIE model fits of observed spectra allow to easily deduce the temperature of a halo, non-equilibrium ionization (NEI) models bear the potential to deduce the thermal history and hence the outflow dynamics, which contains significantly more information, at the expense of running more complicated simulations. If the photon statistics is sufficiently high, XMM-Newton observations permit to  identify emission line complexes, such as O{\sc vii} and O{\sc viii} and others, which in the cases of NGC$\,$3079 and NGC$\,$253 gives poor CIE fits of the whole spectra. Therefore the reduced $\chi^2$ is much better in Fig.~\ref{spectra} (left) for the NEI fit. Our best fit model implies a mass loss rate per unit area of $0.055\, {\rm M}_{\odot}$/kpc$^2$/yr or $3.5\, {\rm M}_{\odot}$/yr for the whole spur region.
\section{\textbf{Summary and Conclusions}}
The disc-halo interaction in star forming galaxies is an essential link in their evolution. Already at moderate star formation rates, a galactic fountain cycle will be established. Part of the outflowing gas will leave the galaxy as a wind, in particular if cosmic rays assist in driving. Recently, Everett et al.\ (2008, see also this volume) have shown, by fitting ROSAT PSPC data of the lower halo of our Milky Way, that also our Galaxy should have a CR driven wind. Although photon statistics will be drastically reduced, external edge-on galaxies are the best laboratories to discriminate between hydrostatic coronae and winds. Since winds modify the halo structure, the dynamics will be modified subsequently. This implies that the thermal and dynamical evolution of the halo have to be calculated simultaneously. In order to fit observed spectra, the synthetic spectra have to be treated equally, i.e.\ they have to be binned into the instrumental channels, and folded through the detector response matrix. Future X-ray missions will combine a large collecting area with a sufficiently high energy resolution (e.g.~by using the micro-calorimeter technology). Further implications of winds are the shallow radial galactic gradient of the diffuse $\gamma$-ray emission  (Breitschwerdt et al.\ 2002, see also Dogiel and Breitschwerdt, this volume), and the acceleration of CRs in galactic wind shocks. Finally, it should be mentioned that the halo magnetic field, which is anchored in the underlying disc, will also promote a global loss of galactic angular momentum.
\\
\textbf{Acknowledgement:} DB thanks his colleagues, Drs.\ Wolfgang Pietsch (MPE) and Andreas Vogler
for permission of reproducing an unpublished figure of a joint
collaboration. VB is the recipient of a DOC-fFORTE fellowship of the Austrian Academy of Sciences at the Institute of Astronomy, Univ.~of Vienna. The XMM-Newton project is supported by the BMBF/DLR and the
MPG.
\end{document}